\documentclass[amsmath,prl,twocolumn,superscriptaddress,nofootinbib]{revtex4}
\usepackage{graphicx}
\usepackage{aas_macros}

\renewcommand{\eqref}[1] {equation $($\ref{#1}$)$}
\newcommand{\comment}[1]{{}}
\def\beq{\begin{equation}}
\def\eeq{\end{equation}}
\def\beqn{\begin{eqnarray}}
\def\eeqn{\end{eqnarray}}

\def\cl{C_{\ell}}

\def\O{\Omega}

\def\Om{\ensuremath{\Omega_{\mathrm{m}}}}

\def\l{\left}
\def\r{\right}

\def\2gcm{\textrm{g cm$^{-2}$}}

\def\modu#1{\l |{#1}\r |}
\def\av#1{\l \langle{#1}\r \rangle}
\def\hmpc{\:{h}^{-1}\mathrm{Mpc}}

\def\H0{\ensuremath{\mathrm{H}_0}}

\def\nn{\nonumber}

\def\fsky{f_{\mathrm{sky}}}

\newcommand{\bmm}[1]{{\mathbf{#1}}}

\newcommand{\bm}[1]{\ensuremath{\mbox{\boldmath $#1$}}}

\def\nl{N_\ell}
\def\sl{S_\ell}
\def\thetaF{\theta_{\mathrm{FWHM}}}
\def\Bell{\bm \ell}
\def\hmpc{ h^{-1} \mathrm{Mpc}}
\def\mnras{Mon. Not. R. Astron. Soc}
\def\physrep{Physics Reports}

\begin{document}
\title{CMB Lensing and the WMAP Cold Spot}
\author{Sudeep Das}
\email{sudeep@astro.princeton.edu}
\affiliation{Department of Astrophysical Sciences, Princeton University, Princeton, NJ 08544, USA}
\author{David N. Spergel}
\email{dns@astro.princeton.edu}
\affiliation{Department of Astrophysical Sciences, Princeton University, Princeton, NJ 08544, USA}
\affiliation{Princeton Center for Theoretical Science, Princeton University, Princeton, NJ 08544, USA}
\affiliation{Astroparticule et Cosmologie APC, 10, rue Alice Domon et 
L\'eonie Duquet, 75205 Paris cedex 13, France}
\date{\today}

\begin{abstract}{Cosmologists have suggested a number of  intriguing hypotheses for the origin of the ``WMAP cold spot'', the coldest extended region seen in the CMB sky, including a very large void and a collapsing texture.  Either hypothesis predicts a distinctive CMB lensing signal.  We show that the upcoming generation of high resolution CMB experiments such as ACT and SPT should be able to detect the signatures of either textures or large voids.  If either signal is detected, it would have profound implications for cosmology.}
\end{abstract}
\maketitle
\section{Introduction}
One of the most intriguing features in the WMAP\footnote{Wilkinson Microwave Anisotropy Probe; http://map.gsfc.nasa.gov/}\citep{bennett.halpern.ea:2003} maps of the microwave sky is the Cold Spot \citep{vielva.martnez-gonzalez.ea:2004,cruz.martnez-gonzalez.ea:2005,cruz.tucci.ea:2006,cruz.cayon.ea:2007}. Under the standard assumption of statistically homogenous Gaussian random fluctuations, the \emph{a posteriori} probability of finding such a feature on the last scattering surface is less than $2\%$ \citep{cruz.martnez-gonzalez.ea:2005,cruz.tucci.ea:2006}. The Cold Spot also appears to have a flat frequency spectrum and is located in a region of low foreground emission,  making it unlikely to be caused by Galactic foregrounds or the Sunyaev-Zel'dovich effect \citep{sunyaev.zeldovich:1970}. This has led some theorists to speculate that the Cold Spot is a secondary effect, generated at some intermediate distance between us and the last scattering surface. One such  model proposes that the Cold Spot may have been caused by the Rees-Sciama effect \citep{rees.sciama:1968} due to an underdense void of comoving radius $\sim 200 h^{-1} \mathrm{Mpc}$ and fractional density contrast $\delta \sim -0.3$ at redshift of $z\lesssim 1$ \citep{inoue.silk:2006,inoue.silk:2007}. Interestingly, \cite{rudnick.brown.ea:2007} reported a detection of an underdense region with similar characteristics  in the distribution of extragalactic radio sources in the NRAO VLA Sky Survey in the direction of the Cold Spot, a claim which has recently been challenged \citep{smith.huterer:2008}. An alternative view \citep{cruz.turok.ea:2007} proposes that the spot was caused by the interaction of the CMB photons with a cosmic texture, a type of topological defect that can give rise to hot and cold spots in the CMB  \citep{turok.spergel:1990}. Bayesian analysis by \cite{cruz.martinez-gonzalez.ea:2008} claims that the texture hypothesis seems to be favored over the void explanation, mainly because such large voids as required by the latter is highly unlikely to form in a $\Lambda$CDM structure formation scenario. Irrespective of whether the Cold Spot was caused by a void or a texture, the CMB photons interacting with such an entity would have been gravitationally deflected. The deflections would lead to a systematic remapping of the primordial CMB anisotropies in and around the Cold Spot. In this brief report, we use simple analytic models for the void and the texture to address the issue of detectability of the gravitational lensing signature of either model, using upcoming high resolution CMB experiments.\par
For calculations presented in this paper, we assume a WMAP 5-year \citep{dunkley.komatsu.ea:2008} flat $\Lambda$CDM cosmology with a total matter density parameter $\Omega_{m} = 0.258$ and a vacuum energy density  $\Omega_{\Lambda} =0.742$. The spectral index of the primordial power spectrum is set to $n_{s} = 0.963$ and the primordial amplitude for curvature perturbations is taken as $A_{s} = 2.41 \times 10^{-9}$ at a pivot scale of $0.002 \hmpc$. The present value of the Hubble parameter is taken as $H_{0} = 72 \mathrm{~km ~s^{-1}~Mpc^{-1}}$. 
\section{Lensing  by the Void}
Gravitational redshift of photons passing through cosmic voids can produce decrements in the observed CMB temperature. This so-called  Rees-Sciama effect \citep{rees.sciama:1968} has been proposed as a possible explanation for the existence of the Cold Spot by  \cite{inoue.silk:2006,inoue.silk:2007}. They assume a compensated spherical underdense region with fractional density contrast  $\delta \sim -0.3$ at $z\sim 1$, and their analysis suggest that the comoving radius of the region required to explain the observed Cold Spot is $\sim 200 h^{-1} \mathrm{Mpc}$. An order of magnitude estimate by \cite{rudnick.brown.ea:2007} for a completely empty void ($\delta=-1$) at $z\lesssim 1$ puts the comoving radius at $\sim 120  \hmpc$. To put these dimensions in perspective, both observations \citep{hoyle.vogeley:2004, patiri.betancort-rijo.ea:2006} and numerical simulations \citep{colberg.sheth.ea:2005,platen..ea:2008} suggest that for $\delta \sim -0.8$ the typical void size tends to be around $\sim 10 \hmpc$. This means that a $\sim 100-200 \hmpc$ void is extremely unlikely to form in the concordance cosmology. Nevertheless, if such a void does exist, its presence should also be apparent through the gravitational deflection of the CMB photons that pass through or near it. \par
Voids, especially the large ones, are seldom spherical and tend to show large axis ratios \citep{shandarin.feldman.ea:2006,platen..ea:2008}. We use  this property to our advantage and model the void responsible for the Cold Spot as a homogeneous cylinder with its axis aligned  along the line of sight. We take its comoving radius to be $r_{v} = 150$ Mpc and its comoving line of sight depth to be $L= 200$ Mpc. The mean redshift of the cylinder is taken to be $\bar z = 0.8$. Under the thin lens approximation \citep{bartelmann.schneider:2001}, the simple  geometry allows us to approximate the cylinder as a disc of surface underdensity $\Delta\Sigma = \delta \av{\Sigma}$ at redshift $\bar z$, where $\av{\Sigma} = \bar{\rho} L/(1+\bar z),$  $\bar\rho$ being the physical background density of the universe at that redshift and $\delta <0 $ denotes the fractional density contrast. This places the cylinder at a comoving distance $D_{L}=2770.3$ Mpc from us and makes its angular radius $R_{V}= 3.1^{\circ}$. To describe a point on the lens plane, we set up a polar coordinate system $(r,\theta)$ on the lens plane, with the origin at the center of the disc. Here we treat $r$ as an angular variable. Using the Gauss's Law for lensing and  the circular symmetry of the problem, we can write the solution for the effective deflection for a CMB photon as $\bm\alpha_{V} = \alpha_{V}(r) \hat r$ where, 
\begin{equation}
\alpha_{V}(r) = 
\begin{cases}
A_{V} r & \text{for  $r < R_{V}$.}\\
 {A_{V}} \frac {R_{V}^2}  r & \text{for $r\ge R_{V}$}
\end{cases}
\label{alphaVoid}
\end{equation}
with 
\beqn
\nn A_{V} &=& \frac{4\pi G}{c^2}\modu{\delta} \av{\Sigma} \frac{D_{LS}}{D_s} \frac{D_L}{(1+\bar z)}\\
\nn & = & \frac32 {\left(\frac{H_0}c\right)}^2 \modu{\delta} \Omega_m L  \frac{D_{LS}}{D_s} {D_L}{(1+\bar z)}\\
& = &  0.01785~\modu{\delta} 
\eeqn 
where in the last step we have substituted the adopted values of the parameters. Here, $D_L$ and  $D_S$ are the comoving distances from the observer to the lens and the source plane (i.e., the last scattering surface).  $D_{LS}$ represents the comoving distance between the lens and the source. \par
Note that the void acts a  diverging lens and the maximum deflection occurs at the edge of the void. For a perfectly empty void $\delta = -1$ the peak deflection has a value of $3.3^{\prime}$, while for a void with moderate underdensity, $\delta \sim  -0.3$ the maximum deflection is $\sim 1^{\prime}$. We would like to point out here that the model of the void we have considered is uncompensated because we have not surrounded it with an overdense shell as is often done when modeling voids. Such a compensated void will have a similar deflection profile inside the void but the deflection angles will rapidly fall to zero at the outer edge of the compensating shell \citep{amendola.frieman.ea:1999}. Since the void itself has a size of $6^{\circ}$ and most of the detection algorithms we will discuss will depend on mapping the CMB in a roughly $8^{\circ}$ square patch around it, the details of the deflection field outside a few degrees of the void will be unimportant for our order of magnitude estimate.
\section{Lensing by the Texture}
An alternate explanation for the anomalous Cold Spot entertains the possibility of a collapsing cosmic texture at $z\sim 6$ that interacted with the CMB photons \citep{cruz.turok.ea:2007}. Textures are cosmic defects that form when a simple Lie group, like SU(2), is completely broken \citep{turok:1989}. The formation and evolution of textures follow a scaling solution in which they collapse and unwind on progressively larger scales. A texture modifies the space-time metric around itself in such a manner that photons that cross it before collapse are redshifted, while those crossing after collapse are blueshifted. Therefore, depending on whether a texture that collapsed at some conformal time $\tau$, was outside or inside the  sphere defined by the currently detected CMB photons at the same time $\tau$, we would observe a cold or a hot spot along the direction of the texture \citep{turok.spergel:1990}. Incidentally, the texture would also produce gravitational deflection of the CMB photons interacting with it. Under the same spherically symmetric scaling approximation as adopted in \cite{cruz.turok.ea:2007}, it can be shown  \citep{durrer.heusler.ea:1992} that to lowest order in the symmetry breaking energy scale, the deflection of a photon trajectory due to a texture can be written as $\bm \beta = - \beta(r)\hat r $, where 
\beq
\beta(r) = 2\sqrt{2} \epsilon \frac{r/R_{T}}{\sqrt{1+ 4 (\frac{r}{R_{T}})^{2}}}.
\eeq
Here $R_{T}$ is the characteristic angular scale of the texture, and is given by 
\beq
\label{RT}
R_{T} = \frac{2\sqrt{2} \kappa (1+z_{T})}{E(z_{T}) \int_{0}^{z_{T}} dz/E(z)},
\eeq
where $E(z) = ({\Om (1+z)^{3}+\O_{\Lambda}})^{1/2}$, $z_{T}$ is the redshift of the texture and $\kappa$ is a fraction of unity. The amplitude of the deflection is set by $\epsilon = 8 \pi^{2} G \eta^{2}$ where $\eta$ is the symmetry-breaking energy scale. Note that in writing the above equation, we have employed a similar co-ordinate system $(r,\theta)$ as we did for the void, on the plane transverse to the line of sight and having its origin at the texture center. The effective deflection angle $\bm \alpha_{T} = - \alpha_{T}(r) \hat r$, by which the  CMB photons are remapped on the sky, is then given by,
\beqn
\nn \alpha_{T}(r) &= & \frac{D_{LS}}{D_{S}} \beta(r)\\
& = & A_{T}  \frac{r}{\sqrt{1+ 4 (\frac{r}{R_{T}})^{2}}},
\label{alphaTexture}
\eeqn
with
\beq
A_{T} =   \frac{2\sqrt{2} \epsilon}{R_{T}} \frac{D_{LS}}{D_{S}}.
\eeq
Unlike the void, the texture acts as a converging lens.\par
Bayesian template fitting for a collapsing texture was performed by \cite{cruz.turok.ea:2007} on the 3-year WMAP data around the Cold Spot using the analytic temperature decrement profile given in \cite{turok.spergel:1990}. Their fit suggests a value of $\epsilon \sim 8\times 10^{-5}$ for the amplitude and $R_{T} \sim 5^{\circ}$ for the scale parameter. The authors argue that their best fit value for $\epsilon$ is biased high due to noise and by performing the same template fitting on several simulated CMB maps with a cold texture spot in each, they find that the true amplitude is close to $4\times 10^{-5}$, consistent with the upper bound, $5\times 10^{-5}$ inferred from the CMB power spectrum \citep{bevis.hindmarsh.ea:2004}. For the lensing template \eqref{alphaTexture}, we therefore adopt the values $\epsilon = 4 \times 10^{-5}$ and $R_{T} = 5^{\circ}$. Texture simulations put the value of $\kappa$ appearing in \eqref{RT} at $\sim 0.1$, which together with the adopted value of $R_{T}$ imply the redshift of the texture to be $z\sim 6$.  This, in turn gives $A_{T} = 5.19 \times 10^{-4}$. Note that the scaling profile in \eqref{alphaTexture} is valid only for comoving distances $r\lesssim R_{T}$ and usually a Gaussian fall-off is assumed beyond this radius. We neglect this detail as we will be interested in detecting the signal on a patch of the order of the size of the Cold Spot. With the values of $A_{T}$ and $R_{T}$ deduced above, the peak deflection near the edge of the Cold Spot will be $\sim 0.1^{\prime}$, more than an order of magnitude smaller than the corresponding value for the void. This can be understood with the following scaling argument. If $M_{<r}$ represents the mass or energy density interior to some radius $r$ in the void or the texture, then the temperature decrement of the CMB photons will be of order the time rate of change of the potential,  $GM_{< r}/(r~t)$, $t$ being a characteristic time scale.  For the void, $t\sim t_{H}$, the Hubble time, whereas for the texture, the characteristic time scale is the light crossing time $t\sim r/c \ll t_{H}$. Therefore, to produce the same temperature decrement, the texture requires less energy density than the void, i.e. $M_{<r}^{texture}<< M_{<r}^{void}$. Since the gravitational deflection  $\alpha \sim G M_{< r}/r$, the deflection due to the texture is expected to be much smaller than that due to the void.
\section{Can CMB Observations Detect Voids and Textures?}
Several ongoing and upcoming CMB experiments have been designed to survey smaller sections of the CMB sky with much higher angular resolutions and sensitivities than ever before.  For example, the Atacama Cosmology Telescope (ACT) \footnote{http://www.physics.princeton.edu/act/} and the South Pole Telescope (SPT) \footnote{http://pole.uchicago.edu/spt/} are designed to map roughly a tenth of the CMB sky at arcminute angular resolution and sensitivities of around 10 $\mu$K per arcminute sky pixel.  In this section, we will estimate the significance with which the void or the texture hypothesis can be confirmed by studying their lensing signatures with a high resolution CMB experiment like ACT. \par
Gravitational lensing caused by massive objects on the line of sight between us and the last scattering surface produces coherent distortions of the small scale features in the CMB, much like the shape distortions of background galaxies due to the lensing by a cluster. The deflection field couples to the large scale gradients in the CMB and correlates the gradients with the small scale features. This property can be used to reconstruct the convergence profile of the lens, a subject that has been studied in detail over recent years \citep{seljak.zaldarriaga:2000,hu.okamoto:2002,hirata.seljak:2003*1,dodelson:2004,vale.amblard.ea:2004,maturi.bartelmann.ea:2005,lewis.king:2006,  hu.dedeo.ea:2007,yoo.zaldarriaga:2008}. In our case, since the lensing template has been already defined by fitting the temperature decrement of the Cold Spot, we can approach the problem in a simpler manner: given a deflection template $\bm\alpha(\bm r) = \alpha(r) \hat r $, we ask how likely is it to be detected by a CMB experiment.\par
We begin by writing the lensed temperature field as,
\beqn
\tilde T(\bm r) &=& T\l(\bmm r + \bm \alpha(r)\r)\nn\\
& \simeq &  T(\bmm r) +  \frac{\partial{T(\bmm r)}}{\partial r}  \alpha(r).
\eeqn
We formulate the problem of detecting the template by introducing a coefficient to the template : $\alpha \rightarrow c\:\alpha$, and constructing the maximum likelihood estimator for $c$. The unlensed temperature field can be written as, 
\beq
\hat T(\bmm r) \simeq \tilde T(\bm r) -c \frac{\partial{\tilde T(\bmm r)}}{\partial r}  \alpha(r),
\eeq
in terms of which, we can write the likelihood function as,
\beq
2\ln {\cal L} = \hat T^{T}(\bmm r) \bmm C^{-1} \hat T(\bmm r)
\eeq
where $\bmm C(\bmm r, \bmm r') = \av{T(\bmm r) T(\bmm r')}$. This gives the following  maximum likelihood estimator for $c$ :
\beq
\hat c =  \frac{\left[\frac{\partial{\tilde T(\bmm r)}}{\partial r} \alpha(r)\right]^T \bmm C^{-1} \tilde T(\bmm r)}{\left[\frac{\partial{\tilde T(\bmm r)}}{\partial r} \alpha(r)\right]^T \bmm C^{-1}\left[\frac{\partial{\tilde T(\bmm r)}}{\partial r} \alpha(r)\right]} .
\eeq
By construction, $\av{c} = 1$; therefore, the signal to noise for detection can be written as, 
\beqn
\label{StoN_real}
\nn \l(\frac{S}{N}\r)^2 &=& \frac 1 {\sigma_c^2} =\frac 12\av{ \frac { {\partial^2{\ln {\cal L}}}}{\partial c^2}} \\
&=& \av{ {\left[\frac{\partial{\tilde T(\bmm r)}}{\partial r} \alpha(r)\right]^T \bmm C^{-1}\left[\frac{\partial{\tilde T(\bmm r)}}{\partial r} \alpha(r)\right]}}.
\eeqn
\begin{figure}[tbp]
\begin{center}
\includegraphics[scale=0.35]{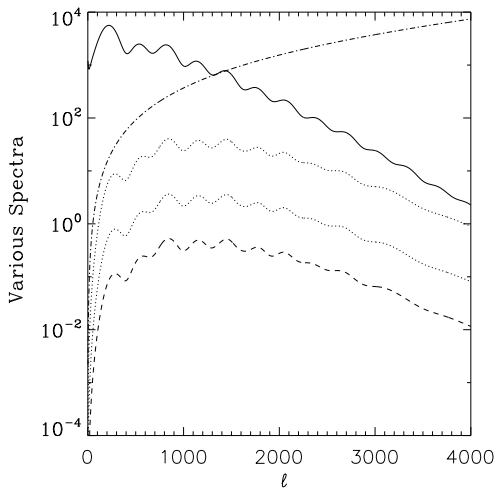}
\caption{Various terms that enter the calculation of the signal to noise \eqref{StoN}. The solid curve represents the CMB power spectrum $\cl$, while the dot-dashed curve represents the instrumental noise for the assumed experimental specifications (see text) and for an exposure time of $16$ minutes. The upper (lower) dotted  curve represents $\sl$ for the void with $\delta = -1$ ($\delta = -0.3$). The dashed line represents $\sl$ for the texture. }
\label{powerSpectra}
\end{center}
\end{figure}
In the Fourier space $\Bell$ conjugate to $\bmm r$, this becomes,  
\beq
\label{StoN}
\l(\frac{S}{N}\r)^2 = \sum_{\ell} \fsky \frac{(2\ell+1)}{2}  \frac {\sl}{\cl+\nl},
\eeq
where we have replaced an integral by a sum and introduced the factor $\fsky$, which  is the fraction of sky area observed, to correct for the fact that all Fourier modes cannot be realized on a finite patch. Here, 
\beq
\label{Signal}
\sl=\int \frac{d^2\bm\ell'}{(2\pi)^2}  \left[\bm\alpha(\bm \ell')\cdot (\bm \ell - \bm \ell') \r]^2 \tilde C_{\modu{\Bell-\Bell'}}
\eeq
and we have used the definition of the power spectrum,
\beq
\av{T^{*}(\Bell) T(\Bell')} = (2\pi)^{2}\cl \delta(\Bell -\Bell').
\eeq
In \eqref{StoN},  $\nl$ is the instrumental noise for the CMB experiment, and is given by
\beq
\nl= 4\pi \fsky \frac{\tau_{e}^{2}}{ t_{\mathrm{obs}}} \exp\left[{\frac{\ell(\ell+1)\thetaF^{2}}{8\ln 2 }}\right]
\eeq
where $\tau_{e}$ is the effective noise-equivalent-temperature (NET) of the detector array (usually expressed in $\mu K \sqrt{s}$), $t_{\mathrm{obs}}$ is the duration of observation, and $\thetaF$ is the full-width-at-half-maximum of the beam, assuming it to be Gaussian.\par
\begin{figure}[tbp]
\begin{center}
\includegraphics[scale=0.35]{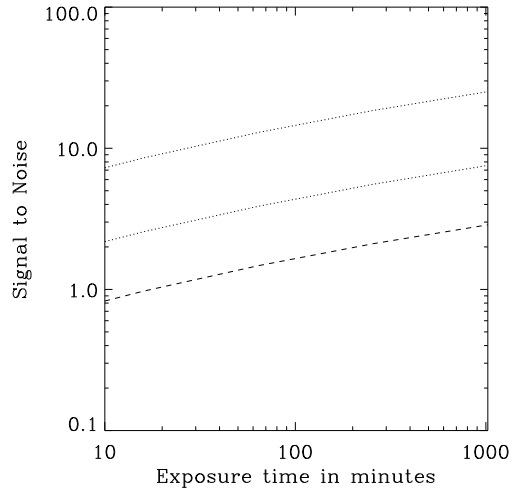}
\caption{Signal to noise for the detection of the lensing template by the experiment described in the text, as a function of the time of exposure of an $8^{\circ}$ square region centered on the Cold Spot. The upper (lower) dotted line corresponds to the case for the void with $\delta = -1$ ($\delta = -0.3$). The dashed line represent the case for the cosmic texture. }
\label{StoNFig}
\end{center}
\end{figure}

In order to evaluate $\sl$, we compute the Fourier transforms of the deflection fields due to the void \eqref{alphaVoid} and the texture \eqref{alphaTexture},
\beq
\bm\alpha(\Bell) = 
\begin{cases}
 - i {\Bell}~  4\pi A_{V} {R_{V}} J_1(\ell R)/{\ell^{3}}& \mathrm{(Void)}\\
 i {\Bell}~  \frac{\pi}{2} A_{T} R_{T}~ e^{-\ell R_{T}/2}(\ell R_{T}+2)/\ell^{3} & \mathrm{ (Texture)} .
\end{cases}
\eeq
Various spectra that enter the calculation of the signal to noise are depicted in Fig.~\ref{powerSpectra}.  For the void we have considered two cases:  a completely empty $\delta=-1$ case, which is a toy model suggested by \citep{rudnick.brown.ea:2007} and the $\delta = -0.3$ case as modeled in detail by \citep{inoue.silk:2006}. As expected, the signal variance $\sl$ for the texture is about an order of magnitude smaller than that of the void with $\delta = -0.3$.  To calculate the signal to noise, we consider a CMB experiment with a $1^{\prime}$ beam and a detector array with effective NET of $\tau_{e} \sim 11~ \mu K ~\sqrt{s}$. We assume that the instrument spends an amount of time $t_{\mathrm{obs}}$ on a $8^{\circ}\times8^{\circ}$ patch containing the Cold Spot, so that $\fsky \sim 1.55\times 10^{-3}$.  Figure~\ref{StoNFig} displays the  signal to noise ratio for the detection of the deflection template as a function of exposure time. Note that although the signal-to-noise per multipole is low (cf. Fig.~\ref{powerSpectra}), so that detection of individual vector modes will be difficult, the total signal-to-noise for the detection that combines the information from all multipoles is high.  It is seen that the $\delta=-1$ void should be readily detectable (or ruled out) at high significance with exposure times of only a few minutes. On the other hand, a significant detection of the texture would require several hours  of integration.  
The calculations above suggest that  both the void and the texture hypotheses can be easily tested by any of the ongoing and upcoming experiments, although realistically, the texture case may need some dedicated allocation of time at the Cold Spot.
\section{Conclusion}
If either a texture or a void is responsible for the WMAP cold spot, then there should be a distinctive lensing signature seen in the CMB.  We have shown that a void would gravitationally lens the CMB photons appreciably so that its presence should be detectable  with arcminute scale CMB experiments. For  a cosmic texture that collapsed at $z\sim 6$,  we find that the gravitational lensing effect on the CMB is more subtle than the void, but should be detectable with longer integration. Together with other indicators, like the power spectrum and the bispectrum \citep{masina.notari:2008} and measurements of the temperature-polarization cross-correlation \cite{cruz.turok.ea:2007}, CMB lensing appears to be a powerful aid in constraining the theories of the WMAP cold spot anomaly. 
\begin{acknowledgements}
We thank Lyman Page, Eiichiro Komatsu, Mark Devlin and Raul Jimenez for useful comments. Sudeep Das thanks APC, Paris for its hospitality during his visit to work on this project.  Das and Spergel acknowledge NASA grant NNX08AH30G and NSF grant 0707731. Sudeep Das thanks Ryan Keisler for pointing out a bug in Fig.~2.
\end{acknowledgements}

\end{document}